\documentclass[a4paper,11pt]{article}
\usepackage{pos}

\usepackage{lineno}
\usepackage[dvipsnames]{xcolor} 
\usepackage{dirtytalk}
\usepackage{ytableau}
\usepackage[dvipsnames]{xcolor}
\usepackage{dsfont}
\usepackage{mathdots}
\usepackage{colortbl}
\usepackage{braket}
\usepackage{amsthm}
\usepackage{amsmath}
\usepackage{float}
\usepackage{bbm, dsfont}
\usepackage[normalem]{ulem}
\usepackage{slashed}
\usepackage{mathtools}
\usepackage{multirow}
\usepackage{cancel}
\usepackage{cleveref}
\usepackage{orcidlink}
\usepackage{enumerate}
\usepackage{fontawesome}
\usepackage{adjustbox}
\usepackage{tcolorbox}
\usepackage{relsize}
\usepackage{csquotes}

\usepackage{tikz}
    \usetikzlibrary{positioning}
    \usetikzlibrary{arrows}
    \usetikzlibrary{shapes}
    \usetikzlibrary{calc}
    \usetikzlibrary{decorations.markings}
     \usetikzlibrary{decorations.pathmorphing}
    \tikzset{snake it/.style={decorate, decoration=snake}}
\def\centerarc[#1](#2)(#3:#4:#5) 
    { \draw[#1] ($(#2)+({#5*cos(#3)},{#5*sin(#3)})$) arc (#3:#4:#5); }
    \usepackage{booktabs}

\newcommand{\bs}[1]{\boldsymbol{#1}}

\title{Canonical differential equations beyond polylogs}

\author[a]{Claude Duhr}
\author*[a]{Sara Maggio}
\author[b,c]{Christoph Nega}
\author[d]{Benjamin Sauer}
\author[c]{Lorenzo Tancredi}
\author[c]{Fabian J. Wagner}

\affiliation[a]{Bethe Center for Theoretical Physics, University of Bonn,\\
  Wegelerstr. 10, D-53115 Bonn, Germany}

\affiliation[b]{Albert-Einstein-Institut, Max-Planck-Institut für Gravitationsphysik,\\
Am Mühlenberg 1, 14476 Potsdam-Golm, Germany}

\affiliation[c]{Technical University of Munich, TUM School of Natural Sciences, Physics Department,\\
James-
Franck-Straße 1, 85748 Garching, Germany}

\affiliation[d]{Institut für Physik, Humboldt-Universität zu Berlin,\\
10099 Berlin, Germany}

\emailAdd{cduhr@uni-bonn.de}
\emailAdd{smaggio@uni-bonn.de}
\emailAdd{christoph.nega@aei.mpg.de}
\emailAdd{benjamin.sauer@hu-berlin.de}
\emailAdd{lorenzo.tancredi@tum.de}
\emailAdd{fabianjohannes.wagner@tum.de}

\abstract{Feynman integrals whose associated geometries extend beyond the Riemann sphere, such as elliptic curves and Calabi–Yau varieties, are increasingly relevant in modern precision calculations. They arise not only in collider cross-section calculations, but also in the post-Minkowskian expansion of gravitational-wave scattering. A powerful approach to compute integrals of this type is via differential equations, particularly when cast in a canonical form, which simplifies their $\varepsilon$-expansion and makes analytic properties manifest. In these proceedings, we will present a method to systematically construct canonical differential equations even for integrals that evaluate beyond multiple polylogarithms. The discussion is kept as light as possible, focusing on the two-loop sunrise integral,
deferring the technical details to the original publications.}

\FullConference{17th International Symposium on Radiative Corrections: Applications of Quantum Field Theory to Phenomenology (RADCOR2025)\\
5-10 October 2025\\
Puri, India\\}

\begin{document}
\begin{flushright}
    BONN-TH-2026-04\\
    TUM-HEP-1591/26\\
    HU-EP-26/06-RTG
    \end{flushright}
\maketitle

\section{Introduction}

Scattering amplitudes are central quantities in Quantum Field Theory. They constitute the backbone for the calculation of most observables at particle
colliders.
Amplitudes are analytic functions of the kinematic invariants, whose non-analyticities are restricted to poles and branch cuts. 
Singularities happen when particle states go on shell, thus  
the singularity structure encodes a large amount of information about the particle content of the theory.
It is therefore crucial that the analytic expression of an amplitude compactly encodes the singularity structure.

In recent years, it has become evident that the computation of an amplitude is streamlined when the tensor structure is factored out, and the problem then reduces to computing scalar \textit{Feynman integrals}.
In what follows, we focus exclusively on scalar integrals.
An extremely powerful technique to compute Feynman integrals is the differential equations method~\cite{Kotikov:1990kg, Remiddi:1997ny, Gehrmann:1999as}, which in turn is based on the existence of \emph{integration-by-parts identities}~\cite{Tkachov:1981wb, Chetyrkin:1981qh} (IBPs) among Feynman integrals. IBPs allow one to reduce all Feynman integrals from a given family to a basis of so-called \emph{master integrals}.  Since Feynman integrals diverge, dimensional regularisation~\cite{tHooft:1972tcz, Bollini:1972ui} can be employed to regularise them. Thus the integrals are functions of the kinematic invariants and the dimensional regulator $\varepsilon = (d-d_0)/2$, with $d_0$ an integer. 

Decades of calculations have shown that the analytic expressions for most one-loop or massless amplitudes can be written as
\begin{equation}
    \mathcal{A}(\bs{s})=\sum_i A_i(\bs{s})\int_{\gamma}\mathrm{d}\log{f_n}(\bs{x},\bs{s})\wedge\ldots\wedge\mathrm{d}\log{f_1}(\bs{x},\bs{s})\,,
    \label{eq:pure_amplitude}
\end{equation}
where the $A_i$ are algebraic functions, dependent on the kinematic invariants $\bs{s}$, that carry the information on the kinematic singularities, whereas the chain of integrations of logarithmic forms constitutes the transcendental part, carrying the information associated with poles and branch cuts. The appearance of logarithmic singularities is expected, as it is a consequence of the locality of the interactions.
Integrals, and amplitudes, with this property are called \textit{pure}~\cite{Arkani-Hamed:2010pyv}, term firstly introduced to describe amplitudes in $\mathcal{N}=4$ super Yang-Mills theory. 

Expressions as in eq.~\eqref{eq:pure_amplitude} are natural when computing one-loop integrals, as they are often associated to a (puctured) Riemann sphere, yielding only simple poles.
However, starting at the two-loop order, Feynman integrals are characterised by higher-dimensional geometries, and $\mathrm{d}\log$-forms are insufficient to span the space of possible solutions. It is then not clear if logarithmic singularities are enough to describe the amplitude.
In these proceedings, we describe how to maintain these desired properties even when the integrals evaluate to functions beyond polylogarithms.

\section{Canonical differential equations and geometries}
\label{sec:2}
Let us start by reviewing the case of Feynman integrals that evaluate to polylogarithms and show what changes when going to a higher-dimensional underlying geometry.

\paragraph{Integrals that evaluate to multiple polylogarithms.}
If the integral evaluates to iterated integrations of rational functions, then the only independent differential forms that may appear in the integrand are of the type
\begin{equation}
    \frac{\mathrm{d}x}{x-a_i}\,,
    \label{eq:diff_forms_polylogs}
\end{equation}
with $a_i$ being the singular points. Thus, the integral can be expressed in terms of iterated $\mathrm{d}\log$ integrations 
\begin{equation}
   J(\bs{s})\sim\int_{\gamma}\mathrm{d}\log{f_n}(\bs{x},\bs{s})\wedge\ldots\wedge\mathrm{d}\log{f_1}(\bs{x},\bs{s})\,\mathcal{G}(\bs{x},\bs{s})^{k\,\varepsilon}\,,
\end{equation}
often evaluating to \textit{multiple polylogarithms} (MPL)~\cite{Kummer, Remiddi:1999ew, Goncharov:1995, Goncharov:1998kja}.
Integrals of this form are not only pure, but are in fact \textit{canonical}~\cite{Henn:2013pwa}, in the sense that they satisfy:
\begin{enumerate}
    \item $\varepsilon$-factorised differential equations: $\mathrm{d}\bs{J}=\varepsilon \bs{A}(\bs{s})\,\bs{J} $;
    \item The entries of the connection $\bs{A}(\bs{s})$ have only simple poles;
    \item The differential forms in $\bs{A}(\bs{s})$ are independent up to total derivatives.
\end{enumerate}
This form is particularly convenient as it allows for a solution in terms of a path ordered exponential
\begin{equation}
    \bs{J}(\bs{s},\varepsilon)=\mathbb{P}\exp\left[\varepsilon\int_{\gamma}\bs{A}(\bs{s'})\right]\,.
\end{equation}
Each order in the Laurent series in $\varepsilon$ can be expressed in terms of Chen iterated integrals~\cite{ChenSymbol} over the differential forms in $\bs{A}(\bs{s})$. 
Moreover, the integral
has uniform transcendental weight (UT)~\cite{Kotikov:2002ab}.

\paragraph{Integrals that evaluate to special functions beyond MPLs.}
For integration domains beyond the Riemann sphere, the differential forms in eq.~\eqref{eq:diff_forms_polylogs} do not suffice to span the space of master integrands. Let us for now specialise to the simplest generalisation of the Riemann sphere: the elliptic curve.
On the elliptic curve, there are still forms that have logarithmic type singularities
\begin{equation}
    \frac{\mathrm{d}x}{(x-a_i)\,y}\,,
\end{equation}
where $a_i$ are the marked points and $y^2=P_{4}(x)$ is the equation defining a quartic elliptic curve.
However, these forms do not suffice to span the cohomology, as a torus has at least two independent cycles beyond those encircling the punctures. The remaining ones are
\begin{equation}
    \frac{\mathrm{d}x}{y}\,\quad \mathrm{and}\quad \frac{\mathrm{d}x}{y} \left(x^2-\frac{s_1\,x}{2}\right)\,,
    \label{eq:elliptic_basis}
\end{equation}
where $s_1$ is the first symmetric polynomial. The forms in eq.~\eqref{eq:elliptic_basis} have no pole and a double pole without residues, respectively.
In the rest of these proceedings we will show that, despite the different pole structure in the differential forms, integrals evaluating to MPLs and beyond MPLs do enjoy similar properties.

\section{Two-loop sunrise integral}
We start by discussing a simple example of a Feynman integral, the \textit{sunrise} integral defined by
\begin{equation}
\label{def:sunrise}
I_{\nu_1,\dots,\nu_{5}}(\bs{s};d) = \int \left(\prod_{j=1}^2\frac{\mathrm d^dk_j}{i\pi^{d/2}}\right)\frac{(k_1 \cdot p)^{-\nu_{4}}(k_2 \cdot p)^{-\nu_{5}}}{(k_1^2-m^2)^{\nu_1}(k_2^2-M^2)^{\nu_2}((k_1-k_2-p)^2-m^2)^{\nu_3}}\,.
\end{equation}
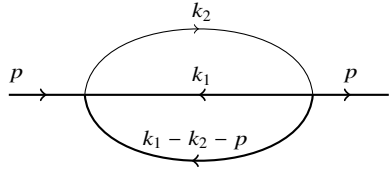
\begin{figure}[!h]
\centering
\begin{tikzpicture}
\coordinate (llinks) at (-2.5,0);
\coordinate (rrechts) at (2.5,0);
\coordinate (links) at (-1.5,0);
\coordinate (rechts) at (1.5,0);
\begin{scope}[very thick,decoration={
    markings,
    mark=at position 0.5 with {\arrow{>}}}
    ] 
\draw [-, thick,postaction={decorate}] (rechts) to [bend right=0]  (links);
\draw [-, thin,postaction={decorate}] (links) to [bend left=85]  (rechts);
\draw [-, thick,postaction={decorate}] (llinks) to [bend right=0]  (links);
\draw [-, thick,postaction={decorate}] (rechts) to [bend right=0]  (rrechts);
\end{scope}
\begin{scope}[very thick,decoration={
    markings,
    mark=at position 0.5 with {\arrow{<}}}
    ]
\draw [-, thick,postaction={decorate}] (links) to  [bend right=85] (rechts);
\end{scope}
\node (d1) at (0,1.1) [font=\scriptsize, text width=.2 cm]{$k_2$};
\node (d2) at (0,0.25) [font=\scriptsize, text width=.2 cm]{$k_1$};
\node (d3) at (0.75,-0.6) [font=\scriptsize, text width=3 cm]{$k_1-k_2-p$};
\node (p1) at (-2.0,.25) [font=\scriptsize, text width=1 cm]{$p$};
\node (p2) at (2.4,.25) [font=\scriptsize, text width=1 cm]{$p$};
\end{tikzpicture}
\caption{The two-loop sunrise graph. The thin and thick lines represent propagators with mass $M$ and $m$, respectively.}
\label{fig:sunrise}
\end{figure}
This is a good example because it allows us to study both the polylogarithmic ($M^2 = 0$) and the elliptic ($M^2 \neq 0$) case. 
It is indeed well known (cf.~refs.~\cite{Sabry, Broadhurst:1987ei, Bauberger:1994by, Bauberger:1994hx, Laporta:2004rb, Bloch:2013tra, Adams:2013nia, Remiddi:2013joa, Adams:2014vja, Adams:2015gva}) that whenever all three propagator masses are non-zero, 
this integral involves functions related to an elliptic curve, while the integral is of polylogarithmic type for $M^2=0$.
The two main questions we seek to address are as follows:
\begin{enumerate}
\item Can we identify master integrals in which the singularity structure is manifest?
\item To what extent are these properties determined by the underlying geometry?
\end{enumerate}

As a starting point, we use the Baikov representation~\cite{Baikov:1996iu, Baikov:1996rk, Frellesvig:2017aai, Frellesvig:2024ymq} to compute maximal cuts and leading singularities~\cite{Cachazo:2008vp}. 
Maximal cuts encode a subset of the analytic structure and, therefore, provide a good starting point for the analysis.
We work close to $d_0=2$ dimensions, which is the most natural (even) 
number of dimensions to analyse this integral.
In the case of the sunrise family, the maximal cut encodes the only non-trivial part of the integral family, because all subtopologies are products of one-loop tadpole integrals. 

Let us start by considering the corner-integral, $I_1 = I_{1,1,1,0,0}(\bs{s};d)$.
Either from a loop-by-loop approach~\cite{Primo:2016ebd,Frellesvig:2017aai} or by parametrising both
loops at once and taking a further residue, we see that the leading singularities of $I_1$ in $d=2$ dimensions are given by a one-fold integral,
\begin{equation}\begin{split}
{\rm LS}\left( I_1 \right) &\,\propto \oint \frac{\mathrm dx_5}{\sqrt{M^2+s+2 x_5} \sqrt{M^2 s-x_5^2}
   \sqrt{4 m^2-M^2-s-2 x_5}}= \oint \frac{\mathrm dx_5}{\sqrt{P_4(x_5)}}\,,
   \label{eq:cutsunrise}
\end{split}\end{equation}
where $P_4(x_5)$ is the quartic polynomial, which, for generic $M^2\neq 0$, defines an elliptic curve: 
\begin{equation}\begin{split}
\mathcal E:\quad y^2 =P_4(x_5) &= (M^2+s+2 x_5)(M^2 s-x_5^2)(4 m^2-M^2-s-2 x_5)\,, \\ 
&= 4(x_5-a_1)(x_5-a_2)(x_5-a_3)(x_5-a_4)\,, \label{eq:P4}
\end{split}
\end{equation}
where the four roots are given by
\begin{align}
    a_1 = - \sqrt{M^2 s}\,, \quad a_2 = - \frac{M^2+s}{2}
    \,, \quad a_3 = \frac{4m^2-M^2-s}{2}\,, \quad a_4 = + \sqrt{M^2 s}\,.
\label{eq:rootsquar}
\end{align}
Thus, as long as the masses are non-zero, the geometry associated to the sunrise integral is the family of elliptic curves defined by eq.~\eqref{eq:P4}. If $M^2=0$, however, the elliptic curve may degenerate. Therefore, we discuss the two cases $M^2 =0$ and $M^2 \neq 0$ separately.

\begin{figure}[h]
    \centering
    \includegraphics[width=0.8\textwidth]{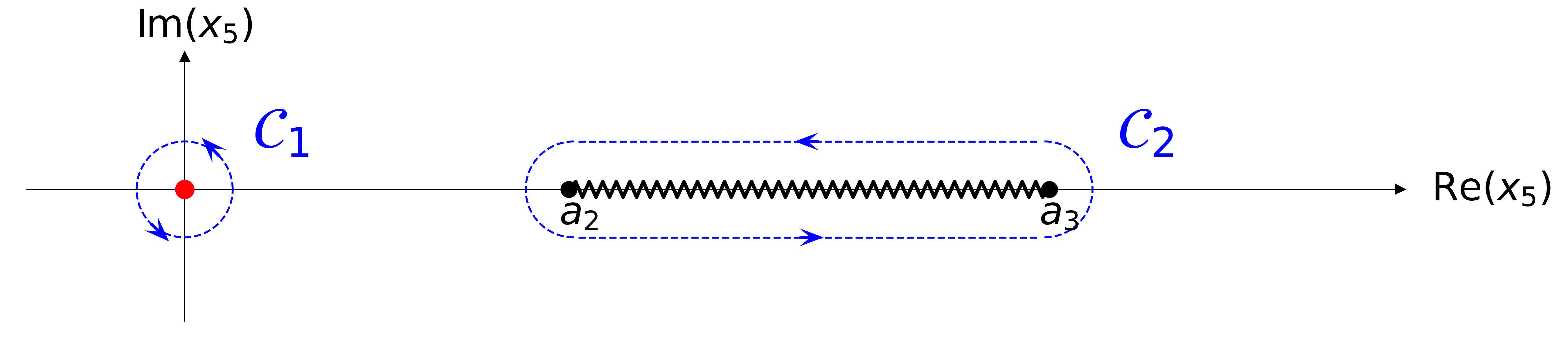}
    \caption{The contours for the two-mass sunrise integral.}
    \label{fig:contours}
\end{figure}

\paragraph{The polylogarithmic case: $M^2 = 0$.}
From eq.~\eqref{eq:P4}, we see that for $M^2=0$ two of the branch points coincide, and the elliptic curve degenerates into two copies of the Riemann sphere connected by the branch cut associated with the
remaining square root. The final geometry is then topologically 
equivalent to a single Riemann sphere. 
By solving the associated IBPs, we see that there are two independent master integrals,
which we can choose as
\begin{equation}
I_1 = I_{1,1,1,0,0}(\bs{s}; d) \textrm{~~~and~~~} I_2 = I_{1,1,1,0,-1}(\bs{s}; d)\,.
\end{equation}
Assuming $m^2>0$ and $s<0$ for definiteness,
we see from eq.~\eqref{eq:cutsunrise} that the leading singularities of the first integral in $d=2$ take the form
\begin{equation}
{\rm LS}\left( I_1 \right)\Big|_{M^2=0} \propto \oint \frac{\mathrm dx_5}{x_5\, \sqrt{s+2 x_5} 
   \sqrt{4 m^2-s-2 x_5}}\,.
   \label{eq:cutsunrise2}
\end{equation}
The integrand of eq.~\eqref{eq:cutsunrise2} has a single pole at $x_5=0$ and a branch cut between the two roots $x_{5,1} = -s/2$ and $x_{5,2} = (4 m^2-s)/2$. Correspondingly, we can consider two independent integration contours: $\mathcal C_1$ is a small circle around the single pole, and $\mathcal C_2$ encircles the branch cut (see~\cref{fig:contours}). 
Since there is no pole at infinity, these two cycles cannot be independent. 
One easily finds that, modulo overall prefactors and with $s\to s+i0$,
\begin{equation}
\begin{aligned}
  \oint_{\mathcal C_1}\frac{\mathrm dx_5}{x_5\, \sqrt{s+2 x_5} 
   \sqrt{4 m^2-s-2 x_5}}&\propto
   \oint_{\mathcal C_2}\frac{\mathrm dx_5}{x_5\, \sqrt{s+2 x_5} 
   \sqrt{4 m^2-s-2 x_5}} 
   \propto\frac{2\pi}{\sqrt{s(s-4m^2)}}\,.
\end{aligned}
\end{equation}
The result is a rational integrand with two poles, whose residues are equal and opposite due to the Global Residue Theorem~\cite{GriffithsHarris}. Thus, we found a candidate with logarithmic singularities.

Let us repeat the same analysis for $I_2$. To obtain its integrand, we just need to
multiply the integrand of $I_1$ by $(k_2 \cdot p) = x_5$. This gives
\begin{equation}
{\rm LS}\left( I_2 \right)\Big|_{M^2=0} \propto \oint \frac{\mathrm dx_5\,}{ \sqrt{s+2 x_5} 
   \sqrt{4 m^2-s-2 x_5}}\,.
   \label{eq:cutsunrise2a}
\end{equation}
The transformation $x_5 \to 1/y_5$ shows that the integrand has a single pole at $x_5 = \infty$ in addition to the branch cut. Again, the residue at the simple pole is related to the integral over $\mathcal{C}_2$, and we get
\begin{align}
\oint_{\mathcal{C}_2} \frac{\mathrm dx_5}{ \sqrt{s+2 x_5} 
   \sqrt{4 m^2-s-2 x_5}} \propto i\pi
   \,,
\end{align}
finding another integral with logarithmic singularities.

To obtain pure integrals, we multiply each integral by the inverse of its leading singularity, so that all algebraic prefactors are removed and the result has purely logarithmic singularities. Our basis of pure master integrals is given by
\begin{equation}
\label{eq:canonical_basis_polylogs}
    J_1 = \sqrt{s(s-4m^2)} I_1 \,, \qquad J_2 = I_2\,.
\end{equation}
The analysis we have carried out so far assumed $\varepsilon=0$. Let us now restore the $\varepsilon$-dependence. This causes no obstruction for integrals of $\mathrm{d}\log$ forms, as is apparent, for example, from the Baikov representation, where the $\varepsilon$-dependence enters as
\begin{equation}
J(\bs{s})\sim\int_{\gamma}\mathrm{d}\log{f_n}(\bs{x},\bs{s})\wedge\ldots\wedge\mathrm{d}\log{f_1}(\bs{x},\bs{s})\,\mathcal{G}(\bs{x},\bs{s})^{k\,\varepsilon}\,.
\end{equation}
The Laurent-expansion around $\varepsilon=0$ generates only additional logarithmic terms and therefore does not spoil the logarithmic singularity structure.

Lastly, we can compute the differential equation, e.g. in $s$, satisfied by the maximal cut of the basis in eq.~\eqref{eq:canonical_basis_polylogs}. It reads
\begin{align} \label{eq:caneqmassless}
   \partial_s\! \begin{pmatrix}
       J_1|_{\textrm{MC}} \\  J_2|_{\textrm{MC}}
   \end{pmatrix}  =
(d-2)   
   \left(
\begin{array}{cc}
 \frac{2 (s-m^2)}{s (s-4 m^2)} & \frac{3}{\sqrt{s(s-4m^2)}} \\
 -\frac{1}{2 \sqrt{s(s-4m^2)}} & -\frac{1}{2 s} \\
\end{array}
\right)\begin{pmatrix}
        J_1|_{\textrm{MC}} \\  J_2|_{\textrm{MC}}
   \end{pmatrix} \,.
\end{align} 
The entries of the differential equation matrix can be identified as dlog-forms. Equation~\eqref{eq:caneqmassless} exactly satisfies our requirements of a canonical differential equation, as discussed in section~\ref{sec:2}.

\paragraph{The elliptic case: $M^2 \neq 0$.}
\label{subsec:ellipticsunrise}
Let us now see how the previous analysis changes if $M^2 \neq 0$. Solving IBPs now exposes three independent master integrals, which fulfil a set of three coupled differential equations.
One possible choice is 
\begin{equation}\begin{split} \label{eq:elliptic_sunrise_basis}
I_1 = I_{1,1,1,0,0}( \bs{s}; d)\,, \qquad I_2 = I_{1,1,1,0,-1}( \bs{s}; d)\,,\qquad
I_3 = I_{1,1,1,0,-2}( \bs{s}; d)\,.
\end{split}\end{equation}
As before, let us study their leading singularities in $d=2$ dimensions. We can write
\begin{equation}
{\rm LS}\left( I_r \right) \propto \oint \mathrm dx_5 \frac{x_5^{r-1}}{\sqrt{P_4(x_5)}}\,,  \qquad r=1,2,3\,,
\end{equation}
with $P_4$ defined in eq.~\eqref{eq:P4}.
Let us repeat the same analysis as in the massless case.

For $r=1$, we reproduce eq.~\eqref{eq:cutsunrise}. In fact, the integrand in eq.~\eqref{eq:cutsunrise} defines a holomorphic differential form on the elliptic curve $\mathcal{E}$ in eq.~\eqref{eq:P4}, and this holomorphic differential is unique up to normalisation. 
Due to the existence of four distinct branch points, there are two \textit{independent} choices of cycles, $\mathcal{C}_1$ and $\mathcal{C}_2$, on $\mathcal{E}$, so that we can compute \emph{two different} leading singularities for $I_1$ in $d=2$ dimensions:
\begin{equation}
\psi_0 = \oint_{\mathcal{C}_1}\frac{\mathrm d x_5}{\sqrt{P_4(x_5)}}\,, \qquad
\psi_1 = \oint_{\mathcal{C}_2} \frac{\mathrm d x_5}{\sqrt{P_4(x_5)}}\,,  \label{eq:persun}
\end{equation}
evaluating to elliptic integrals of the first kind.
With an appropriate choice of cycles, $\psi_0$ is holomorphic around the MUM-point (maximal unipotent monodromy point) $\bs{s}=0$ and $\psi_1$ has a logarithmic singularity. While $\psi_1$ has non-trivial transcendental weight, $\psi_0$ plays a similar role as the square root in eq.~\eqref{eq:canonical_basis_polylogs}. We can then normalise $I_1$ by $\psi_0$ to get a pure integral.

Let us now consider ${\rm LS}\left( I_2 \right)$. 
With the transformation $x_5 \to 1/y_5$, we get
\begin{equation}
{\rm LS}\left( I_2 \right) \propto \oint \mathrm dy_5 \left[ \frac{ 1}{2y_5 } - \frac{s+M^2-2m^2}{4} + \mathcal{O}(y_5)\right] \,.
\label{eq:cutI2}
\end{equation}
This exposes a single pole at $x_5 = \infty$ (or $y_5 = 0$\,). We may correspondingly consider a new independent contour $\mathcal{C}_3$, which encircles the pole at $x_5=\infty$. When evaluated on the contours $\mathcal{C}_1$ or $\mathcal{C}_2$, we obtain an elliptic integral of the third kind. As the residue at $x_5=\infty$ is constant, this
analysis singles out a second master integral, with similar properties as in the polylogarithmic case. 

Finally, consider the leading singularities of the third master integral. 
With the same change of variables as for $I_2$, we find 
\begin{align}
{\rm LS}\left( I_3 \right) \propto \oint  \mathrm dy_5 &\left[ \frac{1}{ 2y_5^2} - \frac{s+M^2-2m^2}{4 y_5}  \,
+ \mathcal{O}(y_5^0)\right] \,. \label{eq:cutI3}
\end{align}
The integrand has both a double and a simple pole at $x_5 = \infty$. 
The simple pole is proportional to the first symmetric polynomial
of the four roots $s_1 = a_1+a_2+a_3+a_4 = s+M^2-2m^2$, and so we may add the appropriate multiple of $I_2$ to cancel the simple pole. As we have seen in eq.~\eqref{eq:elliptic_basis}, this integral is independent from the previous ones, called integral of the second kind. Therefore, provides a third candidate master integral.
However, this integral does not appear to have purely logarithmic singularities.
Moreover, restoring the $\varepsilon$-dependence does not simply amount to adding logarithms. In the presence of higher poles, IBPs mix $\varepsilon$-powers with higher-pole contributions, thereby spoiling the logarithmic expansion. Next we describe how to resolve this issue.

\section{The algorithm}

In this section we summarise the algorithm of refs.~\cite{Gorges:2023zgv, Duhr:2025lbz} applied to the sunrise integral and we briefly discuss its general implementation.
We start by motivating a different choice for the integral with a double pole.
Suppose we found a canonical integral
\begin{equation}
J=c^{(0)}+\varepsilon\sum_k c_k^{(1)}\mathcal{J}_k^{(1)}+\varepsilon^2\sum_k c_k^{(2)}\mathcal{J}_k^{(2)}+\mathcal{O}(\varepsilon^3)\,,
\label{eq:canonical_integral}
\end{equation}
where $\mathcal{J}_k^{(i)}$ are pure integrals and $c^{(i)}$ are such that the integral is UT.
Taking a derivative with respect to any external kinematic
\begin{equation}
\partial J\sim\varepsilon\sum_k A_k \,c_k^{(1)}+\varepsilon^2\sum_k c_k^{(2)}\sum_j A_j\,\mathcal{J}_{k,j}^{(1)}+\mathcal{O}(\varepsilon^3)\,,
\label{eq:derivative_canonical_integral}
\end{equation}
yields still a UT integral, after appropriate multiplication by $\varepsilon^{-1}$. However, eq.~\eqref{eq:derivative_canonical_integral} is not pure anymore, due to the (possibly transcendental) functions $A_i$, acting as leading singularities.
Therefore, $\partial J$ has already some of the properties that are satisfied by a canonical integral.
To understand the leading singularity in this coupled elliptic case, let us introduce the \textit{period matrix} $\bs{W}$.
Briefly, the period matrix pairs the master integrands with the independent cycles and satisfies the same differential equation as the master integrals:
\begin{align}
\mathrm{d}\bs{I}(\bs{s},\varepsilon)&=\bs{\Omega}(\bs{s},\varepsilon)\bs{I}(\bs{s},\varepsilon)\,,\nonumber\\
\mathrm{d}\bs{W}(\bs{s},\varepsilon)&=\bs{\Omega}(\bs{s},\varepsilon)\bs{W}(\bs{s},\varepsilon)\,,
\end{align}
thereby, the entries of $\bs{W}$ provide a minimal linearly-independent basis of solutions. 
Let us now fix $d=d_0$.
In the polylogarithmic case (Riemann-sphere geometry), maximal cuts can be decoupled into $1\times 1$ blocks. 
This is, however, not the case for integrals associated with elliptic curves, where the cohomology of the maximal cut is two-dimensional.
Therefore, when generalising the notion of leading singularities to elliptic integrals, one must consider the full $2\times 2$ period matrix.

Let us now go back to the sunrise family with master integrals in eq.~\eqref{eq:elliptic_sunrise_basis}, and let us focus on the coupled system formed by $\{I_1,I_3\}$. Let us change basis to a derivative basis as motivated above. For $\bs{I}=\{I_1,\partial_s I_1\}$, the period matrix, evaluated for $d=2$, reads
\begin{equation}
\bs{W}=
    \begin{pmatrix}
        \psi_0 & \psi_1  \\
        \partial_s \psi_0 & \partial_s\psi_1
    \end{pmatrix} \,.
\end{equation}
Analysing the series expansion of the entries of $\bs{W}$ around a MUM-point, shows that they are neither pure nor have definite transcendental weight. 
However, it is possible to split $\bs{W}$ into its \textit{semi-simple} and \textit{unipotent} parts
\begin{equation}
\bs{W}  =   
        \underbrace{\begin{pmatrix}
        \psi_0 & 0 \\
        \partial_s \psi_0 & \frac{\Delta}{ \psi_0}
    \end{pmatrix}}_{\bs{W}_{\textrm{ss}}} \cdot
   \underbrace{\begin{pmatrix}
        1 & \frac{\psi_1}{\psi_0} \\
        0 & 1
    \end{pmatrix} }_{\bs{W}_{\textrm{u}}}\,,
\end{equation}
where $\Delta$ is the determinant of $\bs{W}$, an algebraic function.
$\bs{W}_{\textrm{ss}}$ has transcendental weight zero~\cite{Broedel:2018qkq} and can thus be interpreted as the generalisation of the algebraic leading singularities in the polylogarithmic case. $\bs{W}_{\textrm{u}}$ is unipotent, and $\tau=\psi_1/\psi_0$ is a pure function of transcendental weight one\footnote{The weight depends on the normalisation.}, thereby providing a generalisation of UT and pure functions beyond the polylogarithmic case.
Consequently, we multiply the system of $I_1$ and its derivative by the inverse of the semi-simple part of their period matrix to have the correct transcendental weight at $\varepsilon=0$.

Let us now consider the Laurent expansion of the integrals for $\varepsilon\neq0$.
A canonical integral as in eq.~\eqref{eq:canonical_integral} is expected to have as coefficient of $\varepsilon^k$ a pure function of uniform transcendental weight $k$. 
However, while all the entries of $\bs{W}_{\textrm{u}}$ are pure functions, the entries of the second column have different weights. 
This can be remedied by a rescaling by powers of $\varepsilon$ to ensure the expected correlation.
In the case at hand, as also suggested by the analysis of the weight of eq.~\eqref{eq:derivative_canonical_integral}, the goal is reached by rescaling $\partial_s I_1$ by a relative factor of $\varepsilon^{-1}$.

Now that the transcendental weight is realigned and the integrals expand to pure functions, the last ingredient left to ensure is the $\varepsilon$-factorisation of the differential equation. It is possible to check (we refer to ref.~\cite{Duhr:2025lbz} for the details) that $\partial_s \bs{I}=\left(\widetilde{\bs{B}}+\varepsilon \widetilde{\bs{A}}\right)\bs{I}$, with $\widetilde{\bs{B}}$ a nilpotent matrix.
This implies that it can be removed by a unipotent matrix, with the non-constant entries defined by requiring the transformed basis $\bs{J}(\bs{s})=\bs{U}(\bs{s},\varepsilon)\bs{I}(\bs{s})$ to satisfy
\begin{equation}
\mathrm{d}\bs{J}(\bs{s},\varepsilon)=\varepsilon \bs{A}(\bs{s})\bs{J}(\bs{s},\varepsilon)\,.
\end{equation}
As a consequence the functions introduced to $\varepsilon$-factorise ($\varepsilon$-\textit{functions}) satisfy first-order differential equations, which can be solved by series expansions.

Finally, locally close to singular points, we indeed get the properties wished for in section~\ref{sec:2}. Furthermore, the limit $M^2\to 0$ can be taken explicitly and indeed yields $\mathrm{d}\log$ forms.

\paragraph{Beyond elliptic curves.}
Going beyond the elliptic case requires no essential changes, as the algorithm outlined above is not tied to any specific geometry. It can be summarised in two steps:
\begin{enumerate}
\item First step: we start by choosing a \textit{good} initial basis $\bs{I}(\bs{s},\varepsilon)$. We propose to 
choose a basis that is compatible with the mixed Hodge structure associated with the integral at $\varepsilon=0$. In practice, it is obtained by doing an integrand analysis in integer dimensions and choosing integrals aligned with the geometry underlying the maximal cuts at $\varepsilon=0$.
\item Second step: we rotate the initial basis by a sequence of rotations:
\begin{equation}
\bs{J}(\bs{s},\varepsilon)=\bs{U}_{\mathrm{t}}(\bs{s},\varepsilon)\bs{U}_{\varepsilon}(\varepsilon)\bs{U}_{\mathrm{ss}}(\bs{s})\bs{I}(\bs{s},\varepsilon)\,,
\end{equation}
where, $\bs{U}_{\mathrm{ss}}(\bs{s})$ is a matrix whose blocks are inverses of the semi-simple part of the respective period matrices. $\bs{U}_{\varepsilon}$ is needed to re-align the transcendental weight, and finally $\bs{U}_{\mathrm{t}}(\bs{s},\varepsilon)$ is the rotation defined such that the final differential equation matrix is $\varepsilon$-factorised.
\end{enumerate}

\section{Conclusions}
In these proceedings, focusing on the elliptic two-loop sunrise integral, we have described an algorithm to bring integrals with any underlying geometry to canonical form.
So far, it has been applied to a range of cases, including but not limited to the following: elliptic amplitudes and correlators~\cite{Duhr:2024bzt,Forner:2024ojj,Marzucca:2025eak,Becchetti:2025rrz,Coro:2025vgn,Becchetti:2025oyb}, higher-genus curves~\cite{Duhr:2024uid}, black-hole scattering~\cite{Klemm:2024wtd,Driesse:2024feo,Driesse:2026qiz}, and integrals with a multi-scale Calabi-Yau geometry~\cite{Maggio:2025jel,Duhr:2025kkq,Duhr:2025ouy}.
In the future, it will be interesting to gain a deeper understanding of the $\varepsilon$-functions, a first step in this direction has been taken in~\cite{Duhr:2025xyy} and in these proceedings~\cite{Duhr:2026hcs}. It would also be valuable to compare our method with other approaches in the literature~\cite{e-collaboration:2025frv,Bree:2025tug}, to rigorously establish that it always yields a canonical form, and to assess its performance in the numerical evaluation of the special functions that appear in the differential equations.

\acknowledgments

This work was supported in part by the Deutsche Forschungsgemeinschaft (DFG, German Research Foundation) through the Excellence Cluster ORIGINS under Germany’s Excellence Strategy – EXC-2094-390783311 (CN, LT, FW) and through Projektnummer 417533893/GRK2575 ``Rethinking Quantum Field Theory'' (BS), and in part by the European Research Council (ERC) under the European Union’s research and innovation program grant agreements 949279 (ERC Starting Grant HighPHun (CN, LT)), 101043686 (ERC Consolidator Grant LoCoMotive (CD, SM)) and 101167287 (ERC Synergy Grant MaScAmp (CN)).
Views and opinions expressed are, however, those of the author(s) only and do not necessarily reflect those of the European Union or the European Research Council. Neither the European Union nor the granting authority can be held responsible for them.


\bibliographystyle{JHEP}
\bibliography{thebibliography}

@article{Duhr:2024uid,
	archiveprefix = {arXiv},
	author = {Duhr, Claude and Porkert, Franziska and Stawinski, Sven F.},
	doi = {10.1007/JHEP02(2025)014},
	eprint = {2412.02300},
	journal = {JHEP},
	pages = {014},
	primaryclass = {hep-th},
	reportnumber = {BONN-TH-2024-17},
	title = {{Canonical differential equations beyond genus one}},
	volume = {02},
	year = {2025}}

@article{Becchetti:2025rrz,
    author = "Becchetti, Matteo and Coro, Federico and Nega, Christoph and Tancredi, Lorenzo and Wagner, Fabian J.",
    title = "{Analytic two-loop amplitudes for $ q\overline{q}\to \gamma \gamma $ and gg {\textrightarrow} {\ensuremath{\gamma}}{\ensuremath{\gamma}} mediated by a heavy-quark loop}",
    eprint = "2502.00118",
    archivePrefix = "arXiv",
    primaryClass = "hep-ph",
    reportNumber = "TUM-HEP 1554/25",
    doi = "10.1007/JHEP06(2025)033",
    journal = "JHEP",
    volume = "06",
    pages = "033",
    year = "2025"
}

@article{Broadhurst:1987ei,
	author = {Broadhurst, David J.},
	doi = {10.1007/BF01551921},
	journal = {Z. Phys. C},
	pages = {115--124},
	reportnumber = {OUT-4102-21},
	title = {{The Master Two Loop Diagram With Masses}},
	volume = {47},
	year = {1990}}

@article{Bauberger:1994hx,
	archiveprefix = {arXiv},
	author = {Bauberger, S. and Bohm, M.},
	doi = {10.1016/0550-3213(95)00199-3},
	eprint = {hep-ph/9501201},
	journal = {Nucl. Phys. B},
	pages = {25--48},
	reportnumber = {UWITP-06-94},
	title = {{Simple one-dimensional integral representations for two loop selfenergies: The Master diagram}},
	volume = {445},
	year = {1995}}

@article{Adams:2013nia,
	archiveprefix = {arXiv},
	author = {Adams, Luise and Bogner, Christian and Weinzierl, Stefan},
	doi = {10.1063/1.4804996},
	eprint = {1302.7004},
	journal = {J. Math. Phys.},
	pages = {052303},
	primaryclass = {hep-ph},
	title = {{The two-loop sunrise graph with arbitrary masses}},
	volume = {54},
	year = {2013}}

@article{Adams:2015gva,
	archiveprefix = {arXiv},
	author = {Adams, Luise and Bogner, Christian and Weinzierl, Stefan},
	doi = {10.1063/1.4926985},
	eprint = {1504.03255},
	journal = {J. Math. Phys.},
	number = {7},
	pages = {072303},
	primaryclass = {hep-ph},
	title = {{The two-loop sunrise integral around four space-time dimensions and generalisations of the Clausen and Glaisher functions towards the elliptic case}},
	volume = {56},
	year = {2015}}

@article{Remiddi:2013joa,
	archiveprefix = {arXiv},
	author = {Remiddi, Ettore and Tancredi, Lorenzo},
	doi = {10.1016/j.nuclphysb.2014.01.009},
	eprint = {1311.3342},
	journal = {Nucl. Phys. B},
	pages = {343--377},
	primaryclass = {hep-ph},
	reportnumber = {ZU-TH-26-13},
	title = {{Schouten identities for Feynman graph amplitudes; The Master Integrals for the two-loop massive sunrise graph}},
	volume = {880},
	year = {2014}}

@article{Laporta:2004rb,
	archiveprefix = {arXiv},
	author = {Laporta, S. and Remiddi, E.},
	doi = {10.1016/j.nuclphysb.2004.10.044},
	eprint = {hep-ph/0406160},
	journal = {Nucl. Phys. B},
	pages = {349--386},
	reportnumber = {CERN-PH-TH-2004-089},
	title = {{Analytic treatment of the two loop equal mass sunrise graph}},
	volume = {704},
	year = {2005}}

@article{Bauberger:1994by,
	archiveprefix = {arXiv},
	author = {Bauberger, S. and Berends, Frits A. and Bohm, M. and Buza, M.},
	doi = {10.1016/0550-3213(94)00475-T},
	eprint = {hep-ph/9409388},
	journal = {Nucl. Phys. B},
	pages = {383--407},
	reportnumber = {INLO-PUB-9-94, UWITP-2-94},
	title = {{Analytical and numerical methods for massive two loop selfenergy diagrams}},
	volume = {434},
	year = {1995}}

@book{GriffithsHarris,
	author = {P. Griffiths and J. Harris},
	publisher = {Wiley},
	title = {Principles of Algebraic Geometry},
	year = {1978}}

@article{Cachazo:2008vp,
	archiveprefix = {arXiv},
	author = {Cachazo, Freddy},
	eprint = {0803.1988},
	month = {3},
	primaryclass = {hep-th},
	title = {{Sharpening The Leading Singularity}},
	year = {2008}}

@article{Frellesvig:2024ymq,
    author = "Frellesvig, Hjalte",
    title = "{The loop-by-loop Baikov representation {\textemdash} Strategies and implementation}",
    eprint = "2412.01804",
    archivePrefix = "arXiv",
    primaryClass = "hep-th",
    doi = "10.1007/JHEP04(2025)111",
    journal = "JHEP",
    volume = "04",
    pages = "111",
    year = "2025"
}

@article{Adams:2014vja,
	archiveprefix = {arXiv},
	author = {Adams, Luise and Bogner, Christian and Weinzierl, Stefan},
	doi = {10.1063/1.4896563},
	eprint = {1405.5640},
	journal = {J. Math. Phys.},
	number = {10},
	pages = {102301},
	primaryclass = {hep-ph},
	title = {{The two-loop sunrise graph in two space-time dimensions with arbitrary masses in terms of elliptic dilogarithms}},
	volume = {55},
	year = {2014}}

@article{Bollini:1972ui,
	author = {Bollini, C. G. and Giambiagi, J. J.},
	doi = {10.1007/BF02895558},
	journal = {Nuovo Cim. B},
	pages = {20--26},
	title = {{Dimensional Renormalization: The Number of Dimensions as a Regularizing Parameter}},
	volume = {12},
	year = {1972}}

@article{tHooft:1972tcz,
	author = {'t Hooft, Gerard and Veltman, M. J. G.},
	doi = {10.1016/0550-3213(72)90279-9},
	journal = {Nucl. Phys. B},
	pages = {189--213},
	title = {{Regularization and Renormalization of Gauge Fields}},
	volume = {44},
	year = {1972}}

@article{Broedel:2018qkq,
	archiveprefix = {arXiv},
	author = {Broedel, Johannes and Duhr, Claude and Dulat, Falko and Penante, Brenda and Tancredi, Lorenzo},
	doi = {10.1007/JHEP01(2019)023},
	eprint = {1809.10698},
	journal = {JHEP},
	pages = {023},
	primaryclass = {hep-th},
	reportnumber = {CP3-18-58, CERN-TH-2018-211, HU-Mathematik-2018-09, HU-EP-18/29, SLAC-PUB-17336},
	title = {{Elliptic Feynman integrals and pure functions}},
	volume = {01},
	year = {2019}}

@article{Baikov:1996iu,
	archiveprefix = {arXiv},
	author = {Baikov, P. A.},
	doi = {10.1016/S0168-9002(97)00126-5},
	editor = {Werlen, M. and Perret-Gallix, D.},
	eprint = {hep-ph/9611449},
	journal = {Nucl. Instrum. Meth. A},
	pages = {347--349},
	reportnumber = {INP-96-42-449},
	title = {{Explicit solutions of the multiloop integral recurrence relations and its application}},
	volume = {389},
	year = {1997}}

@article{Baikov:1996rk,
	archiveprefix = {arXiv},
	author = {Baikov, P. A.},
	doi = {10.1016/0370-2693(96)00835-0},
	eprint = {hep-ph/9603267},
	journal = {Phys. Lett. B},
	pages = {404--410},
	reportnumber = {INP-96-10-417},
	title = {{Explicit solutions of the three loop vacuum integral recurrence relations}},
	volume = {385},
	year = {1996}}

@article{Frellesvig:2017aai,
	archiveprefix = {arXiv},
	author = {Frellesvig, Hjalte and Papadopoulos, Costas G.},
	doi = {10.1007/JHEP04(2017)083},
	eprint = {1701.07356},
	journal = {JHEP},
	pages = {083},
	primaryclass = {hep-ph},
	title = {{Cuts of Feynman Integrals in Baikov representation}},
	volume = {04},
	year = {2017}}

@article{Primo:2016ebd,
	archiveprefix = {arXiv},
	author = {Primo, Amedeo and Tancredi, Lorenzo},
	doi = {10.1016/j.nuclphysb.2016.12.021},
	eprint = {1610.08397},
	journal = {Nucl. Phys. B},
	pages = {94--116},
	primaryclass = {hep-ph},
	reportnumber = {TTP16-046},
	title = {{On the maximal cut of Feynman integrals and the solution of their differential equations}},
	volume = {916},
	year = {2017}}

@article{Kummer,
	author = {Kummer, E E},
	journal = {J. reine ang. Mathematik},
	pages = {74-90; 193-225; 328-371},
	title = {{\"{U}ber die Transcendenten, welche aus wiederholten Integrationen rationaler Formeln entstehen}},
	volume = {21},
	year = {1840}}

@article{Remiddi:1999ew,
	archiveprefix = {arXiv},
	author = {Remiddi, E. and Vermaseren, J. A. M.},
	doi = {10.1142/S0217751X00000367},
	eprint = {hep-ph/9905237},
	journal = {Int. J. Mod. Phys.},
	pages = {725-754},
	primaryclass = {hep-ph},
	reportnumber = {NIKHEF-99-005, TTP-99-08},
	slaccitation = {CITATION = HEP-PH/9905237;},
	title = {{Harmonic polylogarithms}},
	volume = {A15},
	year = {2000}}

@article{Goncharov:1995,
	archiveprefix = {arXiv},
	author = {Goncharov, A.~B.},
	journal = {Adv.~Math.},
	pages = {197-318},
	title = {{Geometry of configurations, polylogarithms, and motivic cohomology}},
	volume = {114},
	year = {1995}}

@article{Goncharov:1998kja,
	archiveprefix = {arXiv},
	author = {Goncharov, Alexander B.},
	doi = {10.4310/MRL.1998.v5.n4.a7},
	eprint = {1105.2076},
	journal = {Math.Res.Lett.},
	pages = {497-516},
	primaryclass = {math.AG},
	slaccitation = {CITATION = ARXIV:1105.2076;},
	title = {{Multiple polylogarithms, cyclotomy and modular complexes}},
	volume = {5},
	year = {1998}}

@article{ChenSymbol,
	author = {K.~T.~Chen},
	journal = {Bull.\ Amer.\ Math.\ Soc.},
	pages = {831},
	title = {{Iterated path integrals}},
	volume = {83},
	year = {1977}}

@article{Marzucca:2025eak,
    author = "Marzucca, Robin and McLeod, Andrew J. and Nega, Christoph",
    title = "{Two-loop master integrals for mixed QCD-EW corrections to gg{\textrightarrow}H through O({\ensuremath{\epsilon}}2)}",
    eprint = "2501.14435",
    archivePrefix = "arXiv",
    primaryClass = "hep-th",
    reportNumber = "TUM-HEP-1552/25, ZU-TH 02/25",
    doi = "10.1103/d62t-55k2",
    journal = "Phys. Rev. D",
    volume = "112",
    number = "11",
    pages = "113007",
    year = "2025"
}

@article{Sabry,
	author = {Sabry, A.},
	journal = {Nucl. Phys.},
	number = {17},
	pages = {401-430},
	title = {{Fourth order spectral functions for the electron propagator}},
	volume = {33},
	year = {1962}}

@article{Bloch:2013tra,
	archiveprefix = {arXiv},
	author = {Bloch, Spencer and Vanhove, Pierre},
	doi = {10.1016/j.jnt.2014.09.032},
	eprint = {1309.5865},
	journal = {J. Number Theor.},
	pages = {328--364},
	primaryclass = {hep-th},
	title = {{The elliptic dilogarithm for the sunset graph}},
	volume = {148},
	year = {2015}}

@article{Arkani-Hamed:2010pyv,
	archiveprefix = {arXiv},
	author = {Arkani-Hamed, Nima and Bourjaily, Jacob L. and Cachazo, Freddy and Trnka, Jaroslav},
	doi = {10.1007/JHEP06(2012)125},
	eprint = {1012.6032},
	journal = {JHEP},
	pages = {125},
	primaryclass = {hep-th},
	title = {{Local Integrals for Planar Scattering Amplitudes}},
	volume = {06},
	year = {2012}}

@article{Henn:2013pwa,
	archiveprefix = {arXiv},
	author = {Henn, Johannes M.},
	doi = {10.1103/PhysRevLett.110.251601},
	eprint = {1304.1806},
	journal = {Phys. Rev. Lett.},
	pages = {251601},
	primaryclass = {hep-th},
	title = {{Multiloop integrals in dimensional regularization made simple}},
	volume = {110},
	year = {2013}}

@article{Kotikov:1990kg,
	author = {Kotikov, A. V.},
	doi = {10.1016/0370-2693(91)90413-K},
	journal = {Phys. Lett. B},
	pages = {158--164},
	reportnumber = {ITF-90-31E},
	title = {{Differential equations method: New technique for massive Feynman diagrams calculation}},
	volume = {254},
	year = {1991}}

@article{Gehrmann:1999as,
	archiveprefix = {arXiv},
	author = {Gehrmann, T. and Remiddi, E.},
	doi = {10.1016/S0550-3213(00)00223-6},
	eprint = {hep-ph/9912329},
	journal = {Nucl. Phys. B},
	pages = {485--518},
	reportnumber = {TTP-99-49},
	title = {{Differential equations for two-loop four-point functions}},
	volume = {580},
	year = {2000}}

@article{Remiddi:1997ny,
	archiveprefix = {arXiv},
	author = {Remiddi, Ettore},
	doi = {10.1007/BF03185566},
	eprint = {hep-th/9711188},
	journal = {Nuovo Cim. A},
	pages = {1435--1452},
	reportnumber = {DFUB-97-15, DFUB 97-15},
	title = {{Differential equations for Feynman graph amplitudes}},
	volume = {110},
	year = {1997}}

@article{Chetyrkin:1981qh,
	author = {Chetyrkin, K. G. and Tkachov, F. V.},
	doi = {10.1016/0550-3213(81)90199-1},
	journal = {Nucl. Phys. B},
	pages = {159--204},
	title = {{Integration by parts: The algorithm to calculate $\beta$-functions in 4 loops}},
	volume = {192},
	year = {1981}}

@article{Tkachov:1981wb,
	author = {Tkachov, F. V.},
	doi = {10.1016/0370-2693(81)90288-4},
	journal = {Phys. Lett. B},
	pages = {65--68},
	title = {{A theorem on analytical calculability of 4-loop renormalization group functions}},
	volume = {100},
	year = {1981}}

@article{Gorges:2023zgv,
	archiveprefix = {arXiv},
	author = {G\"orges, Lennard and Nega, Christoph and Tancredi, Lorenzo and Wagner, Fabian J.},
	doi = {10.1007/JHEP07(2023)206},
	eprint = {2305.14090},
	journal = {JHEP},
	pages = {206},
	primaryclass = {hep-th},
	title = {{On a procedure to derive \ensuremath{\epsilon}-factorised differential equations beyond polylogarithms}},
	volume = {07},
	year = {2023}}

@article{Becchetti:2025oyb,
    author = "Becchetti, Matteo and Dlapa, Christoph and Zoia, Simone",
    title = "{Canonical differential equations for the elliptic two-loop five-point integral family relevant to tt{\textasciimacron}+jet production at leading color}",
    eprint = "2503.03603",
    archivePrefix = "arXiv",
    primaryClass = "hep-th",
    reportNumber = "DESY 25-029, ZU-TH 13/25",
    doi = "10.1103/zt4w-c1jk",
    journal = "Phys. Rev. D",
    volume = "112",
    number = "3",
    pages = "L031501",
    year = "2025"
}

@article{Duhr:2024bzt,
	archiveprefix = {arXiv},
	author = {Duhr, Claude and Gasparotto, Federico and Nega, Christoph and Tancredi, Lorenzo and Weinzierl, Stefan},
	doi = {10.1007/JHEP11(2024)020},
	eprint = {2408.05154},
	journal = {JHEP},
	pages = {020},
	primaryclass = {hep-th},
	reportnumber = {BONN-TH-2024-12, MITP/24-065, TUM-HEP-1518/24},
	title = {{On the electron self-energy to three loops in QED}},
	volume = {11},
	year = {2024}}

@article{Forner:2024ojj,
    author = "Forner, Felix and Nega, Christoph and Tancredi, Lorenzo",
    title = "{On the photon self-energy to three loops in QED}",
    eprint = "2411.19042",
    archivePrefix = "arXiv",
    primaryClass = "hep-th",
    reportNumber = "TUM-HEP-1539/24",
    doi = "10.1007/JHEP03(2025)148",
    journal = "JHEP",
    volume = "03",
    pages = "148",
    year = "2025"
}

@article{Klemm:2024wtd,
	archiveprefix = {arXiv},
	author = {Klemm, Albrecht and Nega, Christoph and Sauer, Benjamin and Plefka, Jan},
	doi = {10.1103/PhysRevD.109.124046},
	eprint = {2401.07899},
	journal = {Phys. Rev. D},
	number = {12},
	pages = {124046},
	primaryclass = {hep-th},
	reportnumber = {HU-EP-24/02-RTG, TUM-HEP-1492/24},
	title = {{Calabi-Yau periods for black hole scattering in classical general relativity}},
	volume = {109},
	year = {2024}}

@article{Duhr:2025lbz,
    author = "Duhr, Claude and Maggio, Sara and Nega, Christoph and Sauer, Benjamin and Tancredi, Lorenzo and Wagner, Fabian J.",
    title = "{Aspects of canonical differential equations for Calabi-Yau geometries and beyond}",
    eprint = "2503.20655",
    archivePrefix = "arXiv",
    primaryClass = "hep-th",
    reportNumber = "BONN-TH-2025-11, TUM-HEP 1559/25, HU-EP-25/13-RTG",
    doi = "10.1007/JHEP06(2025)128",
    journal = "JHEP",
    volume = "06",
    pages = "128",
    year = "2025"
}

@article{Coro:2025vgn,
    author = "Coro, Federico and Nega, Christoph and Tancredi, Lorenzo and Wagner, Fabian J.",
    title = "{Analytic two-loop amplitudes for di-jet and $\gamma+$jet production mediated by a heavy-quark loop}",
    eprint = "2509.15315",
    archivePrefix = "arXiv",
    primaryClass = "hep-ph",
    reportNumber = "TUM-HEP 1570/25",
    month = "9",
    year = "2025"
}

@article{Driesse:2024feo,
    author = "Driesse, Mathias and Jakobsen, Gustav Uhre and Klemm, Albrecht and Mogull, Gustav and Nega, Christoph and Plefka, Jan and Sauer, Benjamin and Usovitsch, Johann",
    title = "{Emergence of Calabi{\textendash}Yau manifolds in high-precision black-hole scattering}",
    eprint = "2411.11846",
    archivePrefix = "arXiv",
    primaryClass = "hep-th",
    reportNumber = "HU-EP-24/32-RTG, QMUL-PH-24-26, BONN-TH-2024-15, TUM-HEP-1532/24",
    doi = "10.1038/s41586-025-08984-2",
    journal = "Nature",
    volume = "641",
    number = "8063",
    pages = "603--607",
    year = "2025"
}

@article{Maggio:2025jel,
    author = "Maggio, Sara and Sohnle, Yoann",
    title = "{On canonical differential equations for Calabi-Yau multi-scale Feynman integrals}",
    eprint = "2504.17757",
    archivePrefix = "arXiv",
    primaryClass = "hep-th",
    reportNumber = "BONN-TH-2025-16, UUITP--14/25",
    doi = "10.1007/JHEP10(2025)202",
    journal = "JHEP",
    volume = "10",
    pages = "202",
    year = "2025"
}

@article{Duhr:2025kkq,
    author = "Duhr, Claude and Maggio, Sara and Porkert, Franziska and Semper, Cathrin and Stawinski, Sven F.",
    title = "{Three-loop banana integrals with four unequal masses}",
    eprint = "2507.23061",
    archivePrefix = "arXiv",
    primaryClass = "hep-th",
    reportNumber = "BONN-TH-2025-24",
    doi = "10.1007/JHEP12(2025)034",
    journal = "JHEP",
    volume = "12",
    pages = "034",
    year = "2025"
}

@article{Duhr:2025ouy,
    author = "Duhr, Claude and Maggio, Sara",
    title = "{Three-loop banana integrals with three equal masses}",
    eprint = "2511.19245",
    archivePrefix = "arXiv",
    primaryClass = "hep-th",
    reportNumber = "BONN-TH-2025-32",
    month = "11",
    year = "2025"
}

@article{Duhr:2025xyy,
    author = "Duhr, Claude and Maggio, Sara and Porkert, Franziska and Semper, Cathrin and Sohnle, Yoann and Stawinski, Sven F.",
    title = "{Canonical differential equations and intersection matrices}",
    eprint = "2509.17787",
    archivePrefix = "arXiv",
    primaryClass = "hep-th",
    reportNumber = "BONN-TH/2025-30, UUITP--27/25",
    month = "9",
    year = "2025"
}

@article{Bree:2025tug,
    author = "Bree, Iris and others",
    title = "{New algorithms for Feynman integral reduction and $\varepsilon$-factorised differential equations}",
    eprint = "2511.15381",
    archivePrefix = "arXiv",
    primaryClass = "hep-th",
    month = "11",
    year = "2025"
}

@article{Driesse:2026qiz,
    author = "Driesse, Mathias and Jakobsen, Gustav Uhre and Mogull, Gustav and Nega, Christoph and Plefka, Jan and Sauer, Benjamin and Usovitsch, Johann",
    title = "{Conservative Black Hole Scattering at Fifth Post-Minkowskian and Second Self-Force Order}",
    eprint = "2601.16256",
    archivePrefix = "arXiv",
    primaryClass = "hep-th",
    reportNumber = "HU-EP-26/04-RTG",
    month = "1",
    year = "2026"
}

@article{e-collaboration:2025frv,
    author = "Bree, Iris and others",
    collaboration = "{\ensuremath{\varepsilon}}-collaboration",
    title = "{The geometric bookkeeping guide to Feynman integral reduction and $\varepsilon$-factorised differential equations}",
    eprint = "2506.09124",
    archivePrefix = "arXiv",
    primaryClass = "hep-th",
    month = "6",
    year = "2025"
}

@article{Kotikov:2002ab,
    author = "Kotikov, A. V. and Lipatov, L. N.",
    title = "{DGLAP and BFKL equations in the $N=4$ supersymmetric gauge theory}",
    eprint = "hep-ph/0208220",
    archivePrefix = "arXiv",
    doi = "10.1016/S0550-3213(03)00264-5",
    journal = "Nucl. Phys. B",
    volume = "661",
    pages = "19--61",
    year = "2003",
    note = "[Erratum: Nucl.Phys.B 685, 405--407 (2004)]"
}

@inproceedings{Duhr:2026hcs,
    author = "Duhr, Claude and Maggio, Sara and Porkert, Franziska and Semper, Cathrin and Sohnle, Yoann and Stawinski, Sven F.",
    title = "{Intersection theory and canonical differential equations}",
    eprint = "2602.02037",
    archivePrefix = "arXiv",
    primaryClass = "hep-th",
    reportNumber = "BONN-TH/2026-03, UUITP-01/26",
    month = "2",
    year = "2026"
}

\end{document}